# The electronic structure of $EuPd_2Si_2$ in the vicinity of the critical endpoint

O. Fedchenko,[1] Y.-J. Song,[2] O. Tkach,[1] Y. Lytvynenko,[1,3] S. V. Chernov,[4] A. Gloskovskii,[4] C. Schlueter,[4] M. Peters,[5] K. Kliemt,[5] C. Krellner,[5] R. Valentí,[2] G. Schönhense,[1] and H.J. Elmers[1,*]

[1]*Institut für Physik, Johannes Gutenberg-Universität, Staudingerweg 7, D-55128 Mainz, Germany*
[2]*Institut für Theoretische Physik, Goethe-Universität Frankfurt, Max-von-Laue Strasse 1, 60438 Frankfurt am Main, Germany*
[3]*Institute of Magnetism of the NAS and MES of Ukraine, 03142 Kyiv, Ukraine*
[4]*Deutsches Elektronen-Synchrotron DESY, 22607 Hamburg, Germany*
[5]*Physikalisches Institut, Goethe Universität Frankfurt, Max-von-Laue-Strasse 1, 60438 Frankfurt am Main, Germany*

Hard X-ray angle-resolved photoemission spectroscopy reveals significant alterations in the valence band states of $EuPd_2Si_2$ at a temperature $T_V$, where the Eu ions undergo a temperature-induced valence crossover from a magnetic $Eu^{2+}$ state to a low-temperature valence-fluctuating state. The introduction of small amounts of Au on Pd lattice sites and Ge on Si sites, respectively, results in a decrease in $T_V$ and the emergence of an antiferrromagnetic state at low temperatures without valence fluctuations. It has been proposed that the boundary between AFM order and valence crossover represents a first-order phase transition associated with a specific type of second-order critical end point. In this scenario, strong coupling effects between fluctuating charge, spin, and lattice degrees of freedom are to be expected. In the case of $EuPd_2(Si_{1-x}Ge_x)_2$ with $x = 0.13$, which is situated close to the critical end point, a splitting of conduction band states and the emergence of flat bands with one-dimensional character have been observed. A comparison with *ab initio* theory demonstrates a high degree of correlation with experimental findings, particularly in regard to the bands situated in proximity to the critical end point.

## I. INTRODUCTION

The interplay between electronic and structural degrees of freedom in materials with correlated electrons gives rise to interesting phase transitions as evidenced by Gati et al.[1]. The study of these unconventional phase transitions allows for not only the exploration of fundamental aspects of solid-state physics, but also the identification of potential technological applications. In the vicinity of a second-order critical end point (CEP), which terminates a first-order phase transition line, strong fluctuations are anticipated[1]. It is of significant interest to tune a material to a CEP, which can be achieved through the application of mechanical pressure or, alternatively, through the introduction of chemical pressure, namely by partial substitution of atoms with isoelectronic atoms of a different size. In this case, strong-coupling effects between the correlated electrons and the lattice degrees of freedom can be expected, resulting in unconventional phenomena such as critical elasticity[1].

The $EuPd_2Si_2$ system has been utilized as a model for investigating valence transitions induced by temperature or pressure[2–5]. Similar valence transitions have been observed in the related compounds $EuCu_2Si_2$[6,7], and $EuIr_2Si_2$[8–10] under ambient pressure. At elevated pressures, valence transitions have been observed in $EuRh_2Si_2$[11], $EuNi_2Ge_2$[12], and $EuCo_2Ge_2$[13]. Additionally, partial substitutions of atoms within the crystal lattice may elicit a response analogous to that of pressure[14–17].

In their seminal work, Segre et al.[3] presented the first comprehensive phase diagram for $EuPd_2Si_2$ that encompassed both magnetic and valence transitions of this compound. They posited that the phase diagram could serve as a blueprint for understanding the crossover between the highly magnetic divalent $Eu^{2+}$ state at high temperature and the nonmagnetic trivalent $Eu^{3+}$ state at low temperature. It is currently unclear whether the microscopic driving mechanism for the strong intersite interactions responsible for the cooperative valence transition is predominantly electronic or elastic in nature. This is due to the fact that the broken symmetry at the surface presents an obstacle to surface-sensitive photoemission spectroscopy[18–20].

Upon cooling through the temperature of 160 K, a pronounced and continuous valence change from $Eu^{2.8+}$ to $Eu^{2.3+}$ is observed[20–22]. The inflection point of the valence change $V(T)$, designated as $T_V$, serves as a measure of the energy scale associated with the valence change. $EuPd_2Si_2$ is situated on the high-pressure side of the second-order critical endpoint, as evidenced by magnetic measurements[3] and thermodynamic measurements[23]. The valence transition in $EuPd_2Si_2$ can be tuned by external stimuli, including magnetic fields, pressure, and biaxial strain[24–27]. The availability of single crystals of pristine $EuPd_2Si_2$[4,28] and Ge-substituted $EuPd_2(Si_{1-x}Ge_x)_2$[16,17] has facilitated detailed investigation of the valence transition near the critical endpoint[29].

The present study examines the electronic band dispersion is studied in the series $EuPd_2(Si_{1-x}Ge_x)_2$, where the replacement of Si by the isoelectric but larger Ge atom corresponds to $EuPd_2Si_2$ at a negative chemical pressure. In this case, the critical regime can be accessed via the doping concentration.

In this study, hard x-ray angle-resolved photoemission spectroscopy was employed to measure the bulk band structure of $EuPd_2Si_2$ at negative pressure, approaching the critical endpoint of the valence transition. The observation of a flat band accompanied by a splitting of the conduction band indicates the onset of antiferromagnetic order, which is in agreement with the results of density functional calculations.

## II. EXPERIMENTAL

Single crystals of the compounds $EuPd_2(Si_{1-x}Ge_x)_2$ and $Eu(Pd_{1-x}Au_x)Si_2$ were synthesized using the Czochralski method from an Eu-rich levitating melt, as previously described in Ref. 16. To facilitate the pre-orientation of the crystals, Laue diffractograms were employed as the primary characterization tool. Subsequently, the crystals were cleaved using a wire cutter. The single crystals were affixed to the sample holder via epoxy resin, with the (001) plane oriented parallel to the sample holder plate. A stainless steel pin affixed to the surface was utilized for the cleavage process following the introduction of the sample into the ultrahigh vacuum environment. The cleavage was conducted at room temperature under ultrahigh vacuum conditions. The samples were then transferred under vacuum into a He-cooled (25 K) sample stage on a high-precision 6-axis hexapod manipulator of the time-of-flight momentum microscope.

The photoemission experiments were conducted at beamline P22 of the storage ring PETRA III at DESY in Hamburg, Germany[30,31]. The footprint of the photon source is approximately $50 \times 50$ $\mu m^2$. In the 3 keV range, a Si(220) double-crystal monochromator was employed, yielding a total energy resolution of 100 meV. In the 5 keV range, a Si(331) crystal yielded an energy resolution of 150 meV.

One of the most significant advantages of angle-resolved photoelectron spectroscopy in the hard X-ray range is the notable enhancement in the inelastic mean free path of the escaping photoelectrons. Accordingly, the present results represent true bulk properties, which is crucial given that the valency of Eu at the surface of valence fluctuating systems is always divalent, thereby precluding the possibility of studying the valence crossover[8,19,20].

To address the limitations of low cross-section and low signal-to-background ratio in the hard X-ray regime, we employed time-of-flight momentum microscopy[32], a technique that enables the highly efficient acquisition of three-dimensional data sets of the photoelectron intensity $I(E_B, k_x, k_y)$ as a function of binding energy, $E_B = -(E - E_F)$, and momentum $k_x, k_y$. The processing of data to reduce noise limits the resolution of the momentum measurement to 0.08 Å$^{-1}$ for the results presented below.

Photoemission data were acquired for eight hours in each case. The details of the data evaluation procedure are described in Refs.[31,33]. The evaluation and presentation of the band structure data are consistent with those described in Ref.[29].

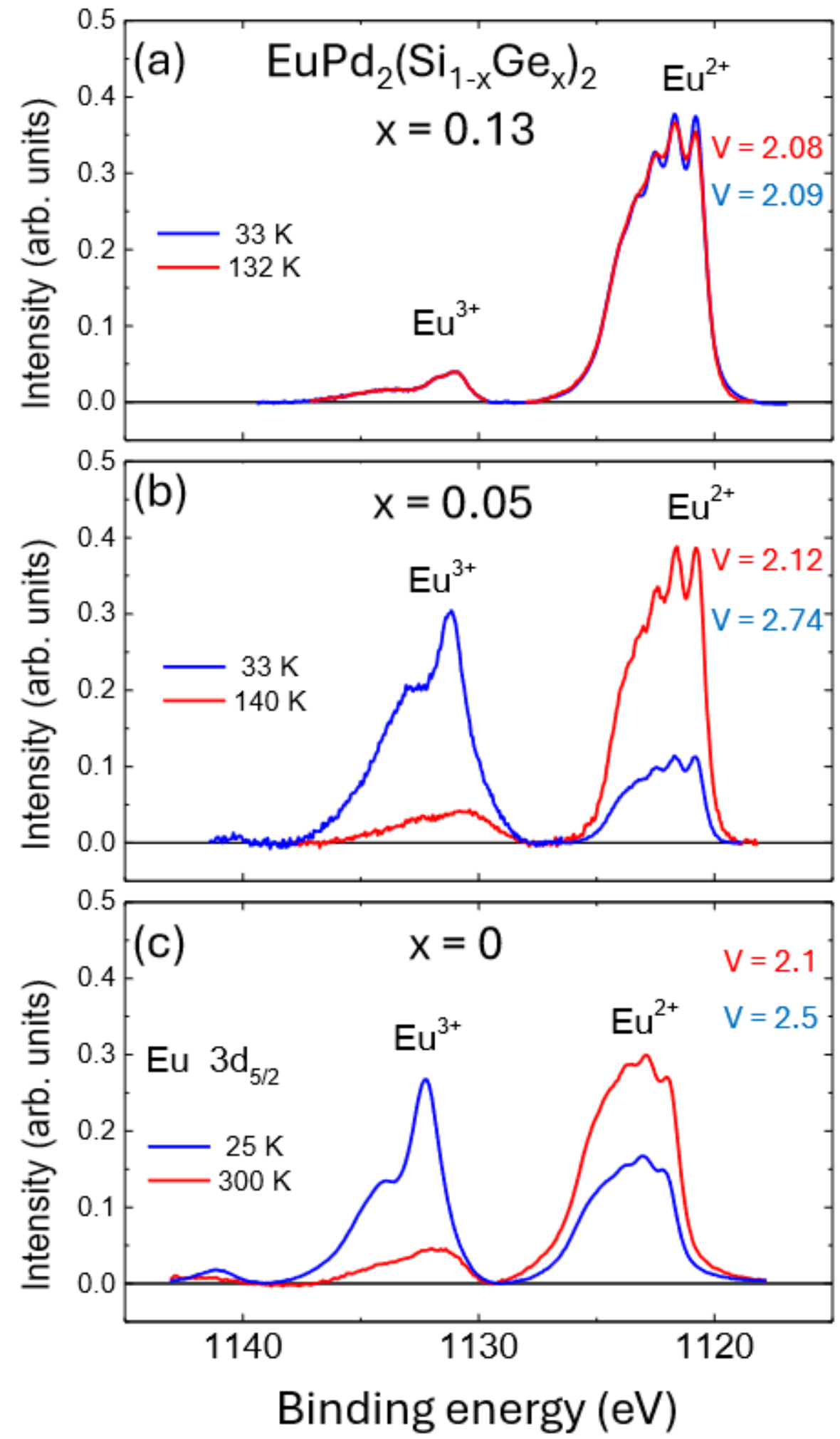

FIG. 1. (a) Temperature dependence of the Eu 3$d$ core level spectra obtained with a photon energy of 3.4 keV at the indicated temperatures for $EuPd_2(Si_{1-x}Ge_x)_2$ single crystals with $x = 0$, 0.05, and 0.13. The Eu valence $v$ is calculated from the ratio of the areas of the corresponding 3$d$ core level peaks. The specified binding energies are referred to the Fermi level. Data for $x = 0$ are taken from Ref.[29].

## III. EXPERIMENTAL RESULTS FOR $EuPd_2(Si_{1-x}Ge_x)_2$

The valence number $V$ of the Eu ions is determined at temperatures above and below the valence transition using hard x-ray photoelectron spectroscopy (HAXPES). This is conducted at the Eu 3$d$ core levels at a photon energy of 3.4 keV. The high kinetic energy of photoemitted electrons ensures bulk sensitivity[34]. Figure 1 compares the experimental results for $EuPd_2(Si_{1-x}Ge_x)_2$ with $x = 0$, 0.05, and 0.13. The spin-orbit interaction results in the splitting of the Eu $3d$ spectrum into two components, Eu $3d_{5/2}$ and an Eu$3d_{3/2}$, component with a splitting energy of 30 eV. In this study, we present the results for the Eu $3d_{5/2}$ component exclusively. The Eu $3d_{5/2}$ component is further split by a chemical shift contingent on the valence state, which is either $Eu^{2+}$ or $Eu^{3+}$ with a splitting of 10 eV. The splitting enables the determination of the mean valence from the corresponding peak areas[35].

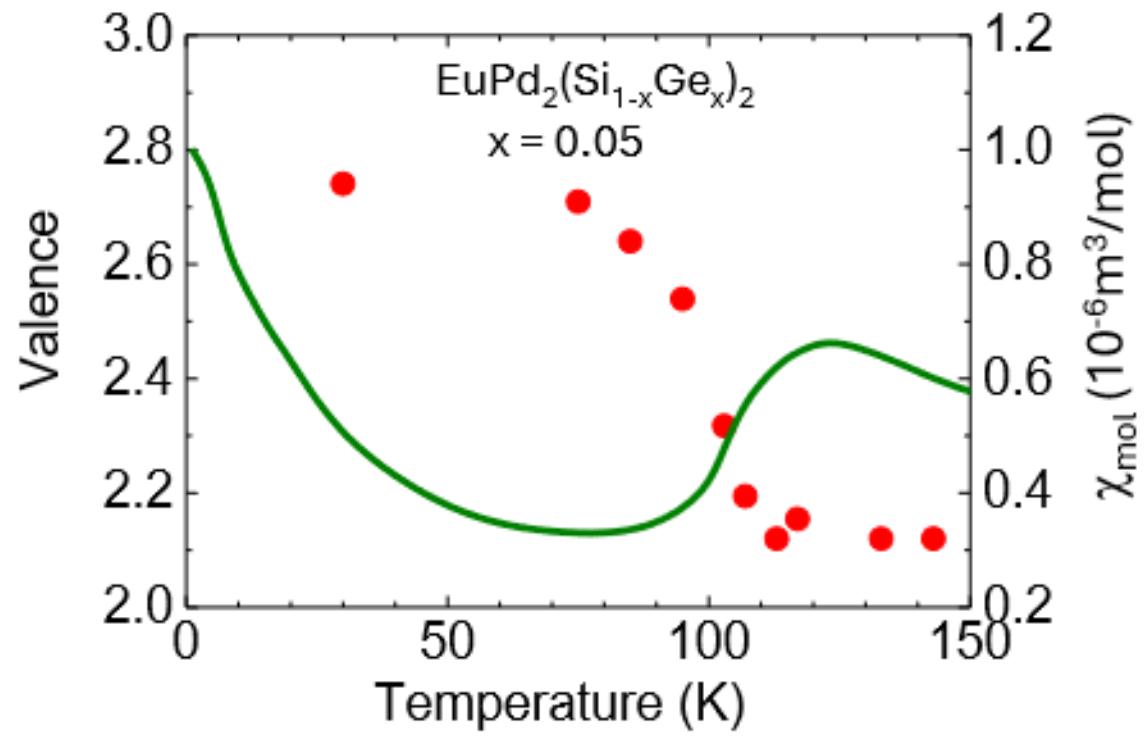

FIG. 2. Temperature dependence of Eu valence for $EuPd_2(Si_{1-x}Ge_x)_2$ with $x = 0.05$ as determined from the Eu HAXPES spectra (red points, left scale). Molar susceptibility as a function of temperature, indicating a similar transition temperature (green curve, right scale).

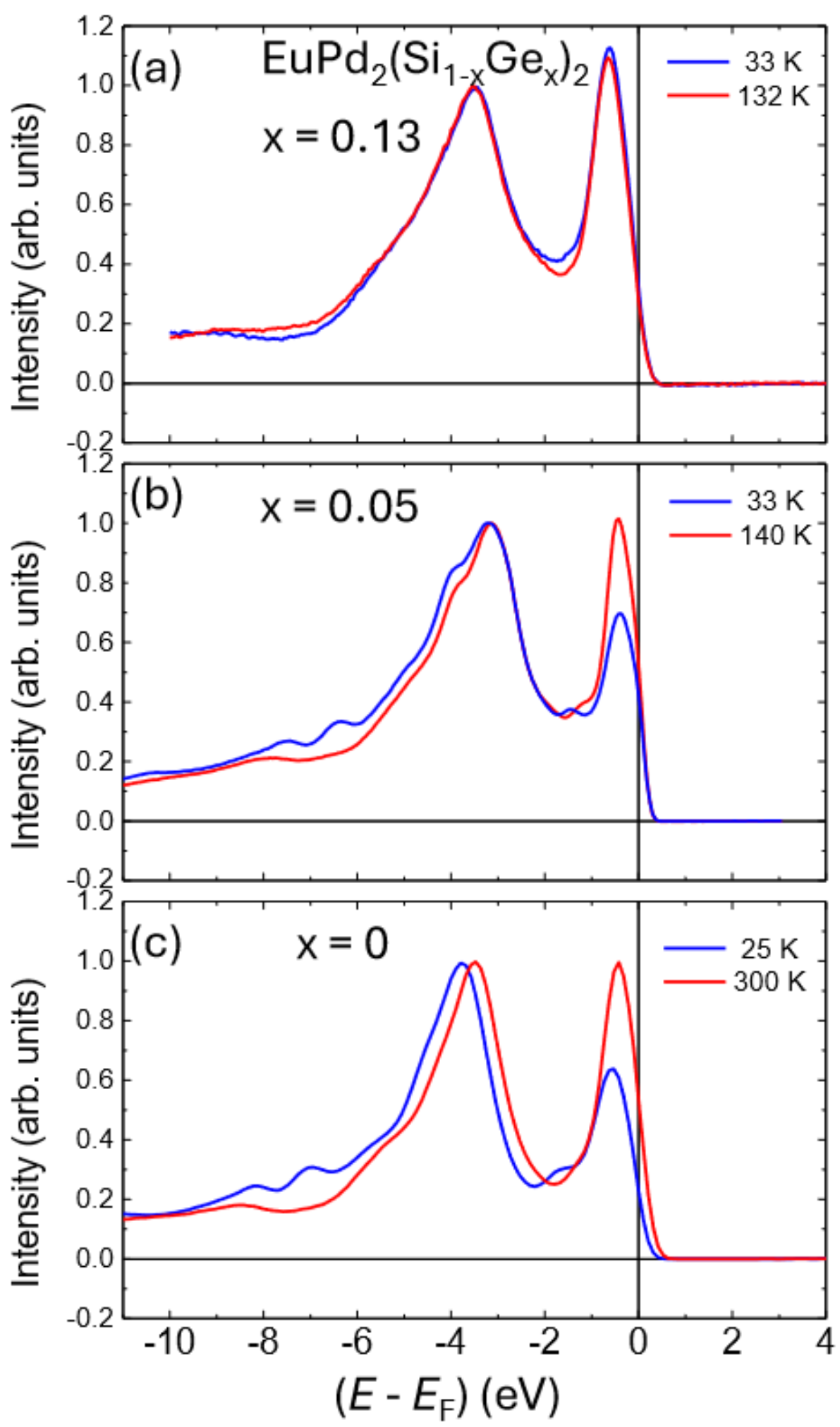

FIG. 3. Valence band energy distribution curves for the indicated temperatures and samples. The photoemission intensity was integrated over a planar section of the full Brillouin zone.

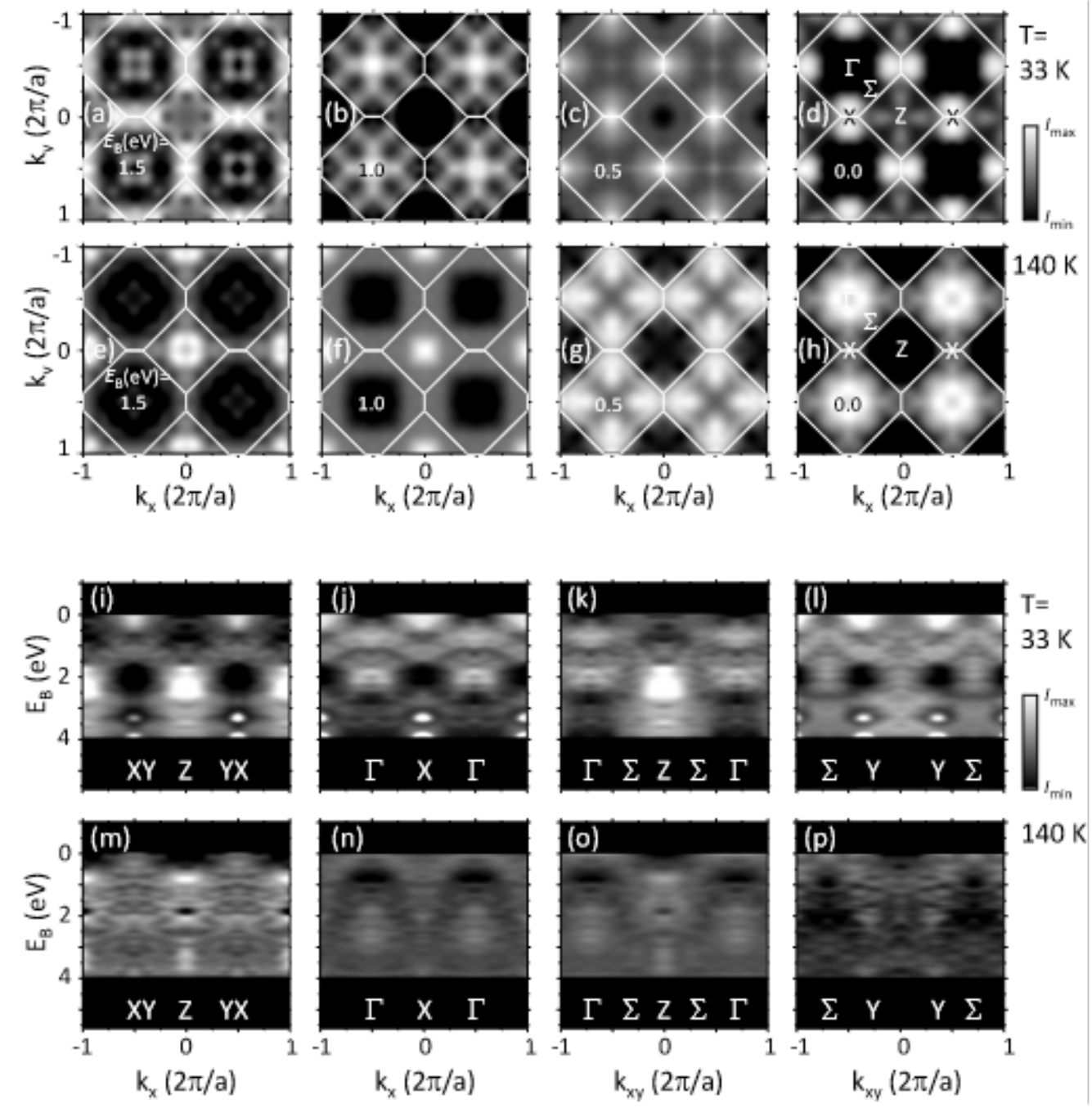

FIG. 4. (a-h) $EuPd_2(Si_{1-x}Ge_x)_2$ with $x = 0.05$. Constant energy maps of the photoemission intensity $I(E_B, k_x, k_y)$ for the indicated binding energies $E_B$ measured at 33 K (a-d) and 140 K (e-h). The photon energy is 3.4 keV. The photoemission intensity has been symmetrized according to the crystal symmetry. (i-p) Binding energy versus parallel momentum sections of the photoemission data array along the indicated high symmetry directions measured at 33 K (i-l) and 140 K (m-p). The photoemission intensity is normalized on the valence band energy distribution curve shown in Fig. 3(b) and color coded on the indicated linear black-white scale.

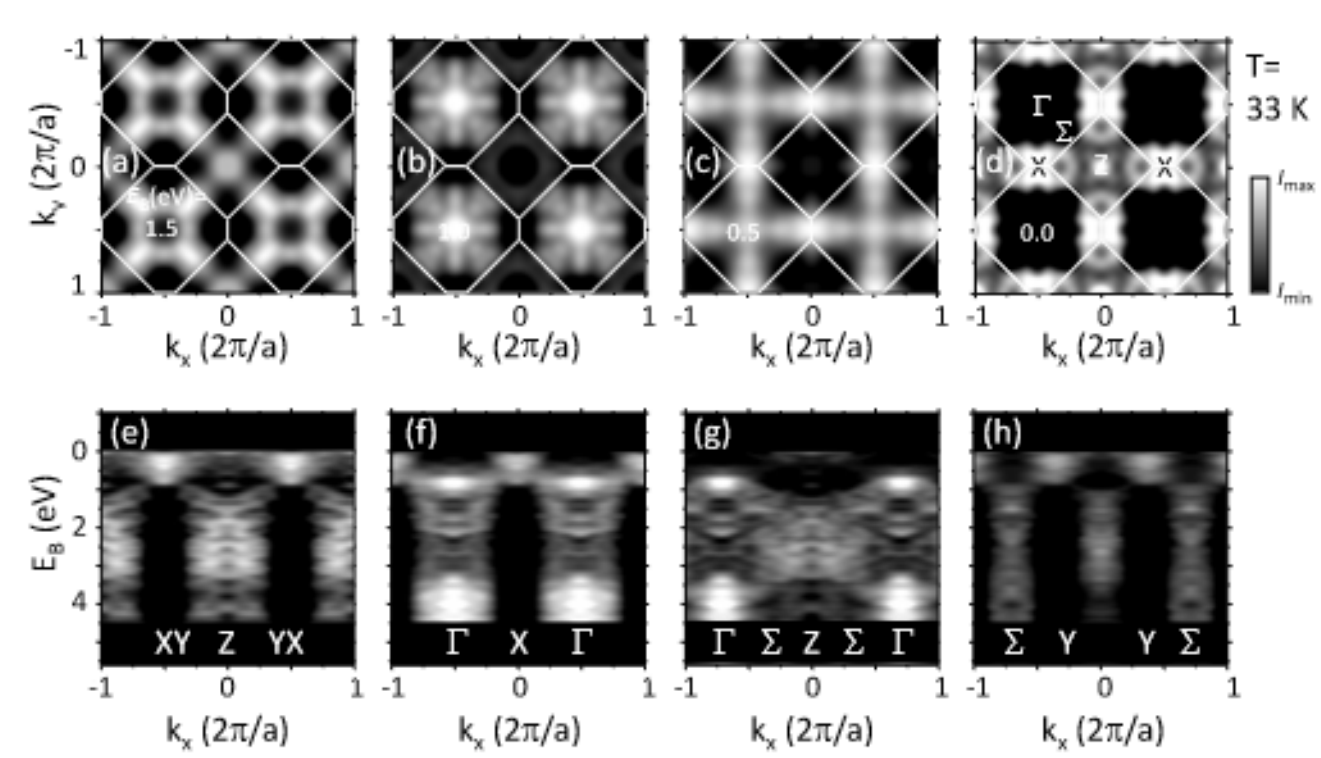

FIG. 5. (a-h) $EuPd_2(Si_{1-x}Ge_x)_2$ with $x = 0.13$. Constant energy maps of the photoemission intensity $I(E_B, k_x, k_y)$ for the indicated binding energies $E_B$ measured at 33 K (a-d). The photon energy is 4.9 keV. The photoemission intensity has been symmetrized according to the crystal symmetry. (i-p) Binding energy versus parallel momentum sections of the photoemission data array along the indicated high symmetry directions measured at 33 K (e-h). The photoemission intensity is normalized on the valence band energy distribution curve shown in Fig. 3(a) and color coded on the indicated linear black-white scale.

A notable shift in the ratio between the $Eu^{2+}$ and $Eu^{3+}$ components is observed for the cases of $x = 0$ and 0.05, indicative of a valence transition. For $x = 0$, an evaluation of the peak area ratio yields valence number values of $V = 2.5$ at low temperature $V = 2.1$ at high temperature. In the case of partial substitution of Si by Ge with $x = 0.05$, the high-temperature valence number exhibited minimal change with a value of $V = 2.12$. However, at low temperature, the value increased to $V = 2.74$, indicating an even more pronounced valence transition compared to the pristine sample. For higher Ge concentration of $x = 0.13$, the valence numbers for high and low temperature are nearly identical, at $V = 2.08$ and 2.09, respectively. In this case, the valence transition is not observed.

Figure 2 illustrates the temperature dependence of the valence number for the case of $x = 0.05$, as determined from the HAXPES spectra. The inclination point indicates the valence transition temperature, which is determined to be 100 K. This value is in agreement with the inclination point of the susceptibility curve. However, it should be noted that minor discrepancies may be attributed to the differing experimental setups employed. The transition is not abrupt, but rather occurs within a temperature range between 90 K and 110 K.

The valence band photoemission intensity measurements, as illustrated in Fig. 3, demonstrate the density of states at both low and high temperatures. For the sake of comparison, the spectra have been normalized to the maximum at 3.5 eV binding energy. The sharp peak situated in close proximity of the Fermi level can be attributed to the partially filled $Eu^{2+}$ $4f$ states. The observed increase in this peak for $x = 0$ and 0.05 in the high temperature phase is indicative of an increase in the $Eu^{2+}/Eu^{3+}$ ratio. The $Eu^{3+}$ $4f$ states are responsible for the two minor peaks observed at 2 and 8 eV binding energy at low temperature, which are no longer present at high temperature. The valence spectra also corroborate the existence of an almost pure magnetic $Eu^{2+}$ state for temperatures exceeding the valence transition temperature ($T > T_V$) and a pronounced redistribution of the peak heights upon cooling below $T_V$. These findings align with the observations made on polycrystalline $EuPd_2Si_2$ samples[19].

In contrast to the low Ge concentrations $x = 0$ and 0.05, the higher concentration of $x = 0.13$ suppresses the valence transition. There is no significant temperature dependence in this case, as illustrated in Fig. 3(a). The sharp peak near the Fermi level associated with the $Eu^{2+}$ $4f$ states is the highest of all measured compounds.

The results of hard x-ray momentum microscopy yield a data array of the photoemission intensity $I(E_B, k_x, k_y)$ that is simultaneously measured as a function of binding energy $E_B$ and parallel momentum $k_x, k_y$. The photoemission intensity resulting from phonon scattering is considerable and superimposes the intensity from direct photoemission processes. This component can be approximated by the matrix-element-weighted density of states, which is independent of the parallel momentum [27] and depicted as the valence band energy distribution curves integrated over a planar section of the full Brillouin zone (see Fig. 3).

Figures 4(a-h) illustrate the alteration in the constant energy intensity maps, $I(E_B, k_x, k_y)$, between temperatures of 33 K and 140 K for the specific case of $x = 0.05$. At 33 K, the section at the Fermi level [Fig. 4(d)] in the Γ-Σ-X plane exhibits ellipses centered on the X-points, which are the brightest features, with the long axis oriented along the Γ-X direction. At room temperature, the high photoemission intensity at the X-points is no longer visible [Fig. 4](h)]. Instead, a high intensity now occurs at the Γ-points. The opposite intensity behavior occurs at $E_B = 1$ eV. Here, a high intensity occurs at the Γ-points at 33 K, while the intensity has vanished at 140 K.

In order to emphasize the band dispersion, we divided the momentum-resolved intensity, $I(E_B, k_x, k_y)$, by the integrated intensity within the Brillouin zone, as shown in Fig.3(b), and subtracted the momentum-independent background intensity. It should be noted that due to this normalization the high intensity of the flat (dispersionless) Eu $4f$ and Pd $4d$ bands at $E_B = 0.5$ and 3.5 eV do not appear in this representation. Following this data processing, the band dispersion of the conduction bands is clearly visible [see Figs. 4(i)-4(p)]. It should be noted, however, that the energy resolution is insufficient to observe any hybridization of the localized Eu $4f$ with other valence states, which has been observed with higher energy resolution[36].

At 33 K, the corresponding intensity distribution demonstrates the presence of electron bands in proximity to the X-points, exhibiting electron-like, parabolic dispersion characteristics and a maximum binding energy of 0.5 eV [for details, please refer to Fig. 4(i,j,l)]. The effective mass of this band is equal to the mass of the electron in the vacuum, $m/m_0 = 1$. At 140 K, the band has shifted to a above the Fermi level and is no longer discernible [see Fig. 4(m.o,p)]. Given that this change is accompanied by a valence change from trivalent $Eu^{3+}$ to divalent $Eu^{2+}$, it seems plausible to suggest that the electron occupying the free-electron-like state at low temperature has become localized at 140 K. This would explain the increase in the peak at $E_B = 0.5$ eV, as illustrated in Fig. 3(b).

A pronounced band dispersion is observed along the Y-Σ direction [Fig. 4(l)], with a mean binding energy of $E_B = 1$ eV and an energy width of 1 eV. The maxima are located at the Σ-points, while the minima are situated at the Y-points. At 140 K, this band has shifted to higher binding energies and has become partly unoccupied near the Σ-points.

These results are, to some extent, similar to those previously published for $x = 0$[29]. The replacement of 5% of the Si atoms by Ge atoms results in a significant shift in transition temperature, while the dispersion behavior and its behavior above and below the transition temperature remain largely unchanged. In this sense, the substitution can be considered to act as a negative pressure, which is in accordance with expectations.

We now consider the case of $x = 0.13$, which is located in close proximity to the critical end point. As previously demonstrated by the XPS data, the valence transition does not occur between 27 K and room temperature [Fig. 1(a)]. Furthermore, the integrated valence band intensity shows no notable variation within this temperature range. Fig. 5 illustrates the corresponding momentum dependence of the valence band states at 33 K. The data has been processed in a manner analogous to that employed for the data presented in Fig. 4.

As illustrated in Fig. 5(d), the constant-energy section of the Fermi surface for the function $I(E_B, k_x, k_y)$ reveals the presence of rectangular-shaped regions of high intensity centered at the X-points. The intensity minima are observed at the Γ-points. At a binding energy of $E_B = 1$ eV, high intensity is observed at the Γ-points, while the intensity is lower at the X-points. This is noteworthy as these bands bear resemblance to the result for $x = 0.05$ discussed previously and $x = 0$[29], despite the valence state being that of the high temperature phase of pristine $EuPd_2Si_2$.

In contrast to $EuPd_2(Si_{1-x}Ge_x)_2$ with $x = 0$ and $x = 0.05$, the constant energy section at $E_B = 0.05$ eV demonstrates a notable absence of dispersion along the Γ-X directions [Fig. 5(c)], forming narrow lines of high intensity. This is indicative of quasi-one-dimensional states constrained along the Γ-X high-symmetry directions, which may be attributed to the onset of strong fluctuations in the vicinity of the critical end point.

A second noteworthy observation is the apparent splitting of the electron-like bands in the vicinity of the X-points, with the apex at $E_B = 0.2$ and 0.5 eV. This splitting may be attributed to the onset of antiferromagnetism, which is evidenced by the corresponding exchange splitting of 0.3 eV.

## IV. EXPERIMENTAL RESULTS FOR $Eu(Pd_{1-x}Au_x)Si_2$

The partial substitution of Pd by the larger Au atoms causes a negative pressure, too. In fact, the $Eu(Pd_{1-x}Au_x)Si_2$ system was the first system to study the influence of doping on the valence transition[3]. A possible coexistence of both mixed valence and magnetic order has been detected. Below a critical concentration of $x = 0.175$ the system behaved mixed valent, above this concentration (up to $x = 0.25$) the system ordered magnetically keeping a weak mixed valent character[3]. The $Eu(Pd_{1-x}Au_x)Si_2$ system showed an increasing instability of the magnetic $Eu^{2+}$ ground state configuration from $x = 1$ to 0.18[37].

A comparison of the high (132 K) and low (27 K) temperature spectra of $Eu(Pd_{1-x}Au_x)Si_2$ with $x = 0.05$ shows a significant change in the ratio between the $Eu^{2+}$ and $Eu^{3+}$ components [Fig. 6]. An evaluation of the peak area ratio results in values for the valence number $V = 2.35$ for $T > T_V$, corresponding to a prevailing $Eu^{2+}$ ionization state, and an increase to $V = 2.5$ at low temperature, indicating a mixed $Eu^{2+}/Eu^{3+}$ valence state.

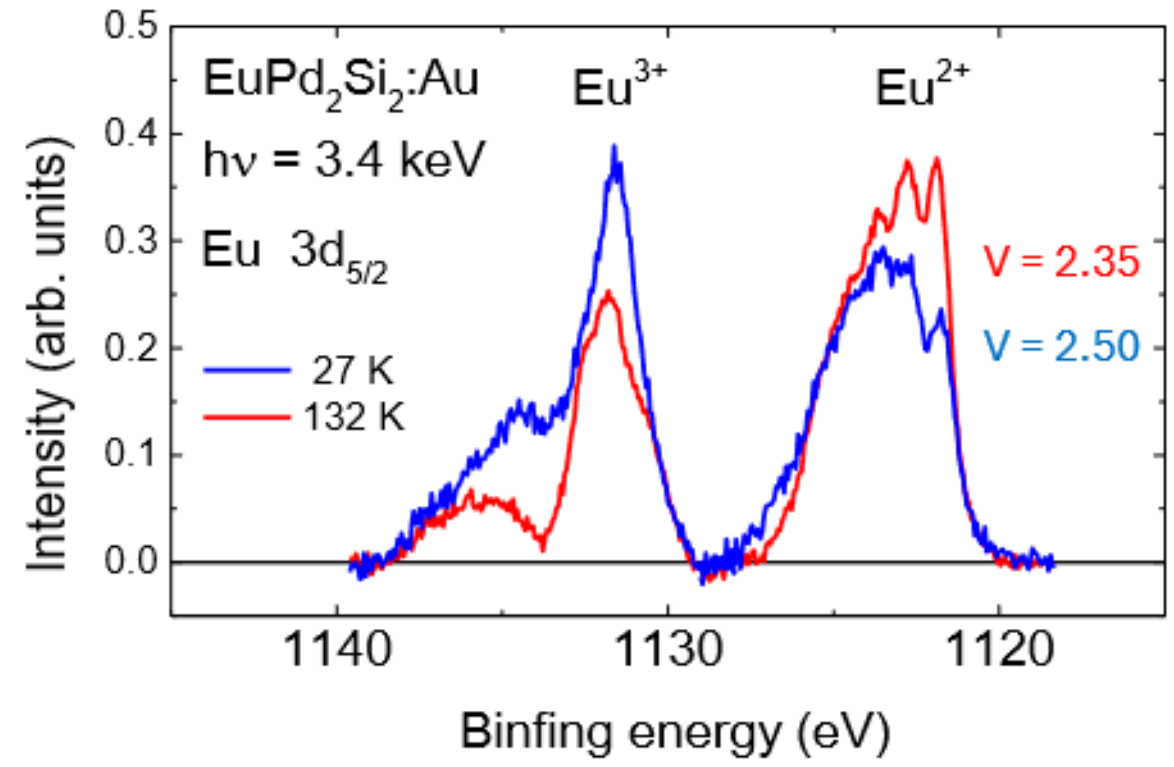

FIG. 6. (a) Temperature dependence of the Eu $3d$ core level spectra obtained with a photon energy of 3.4 keV at the indicated temperatures for a $Eu(Pd_{1-x}Au_x)Si_2$ single crystals with $x = 0.03$. The Eu valence $v$ is calculated from the ratio of the areas of the corresponding $3d$ core level peaks. The specified binding energies are referred to the Fermi level.

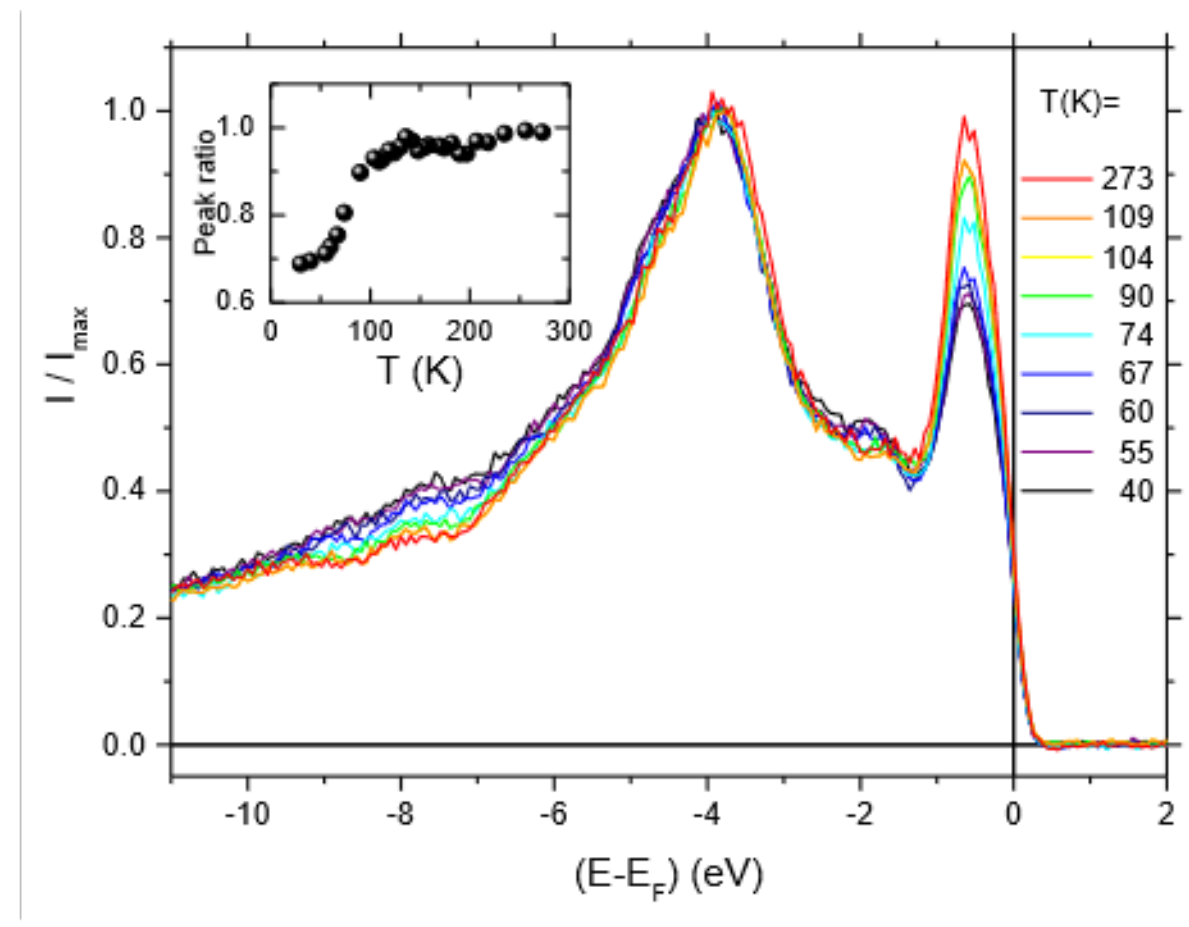

FIG. 7. Valence band energy distribution curves for the indicated temperatures for $Eu(Pd_{1-x}Au_x)Si_2$ with $x = 0.03$. The photoemission intensity was integrated over a planar section of the full Brillouin zone. The inset shows the temperature dependence of the peak ratio, indicating the transition temperature $T_v = 75$ K.

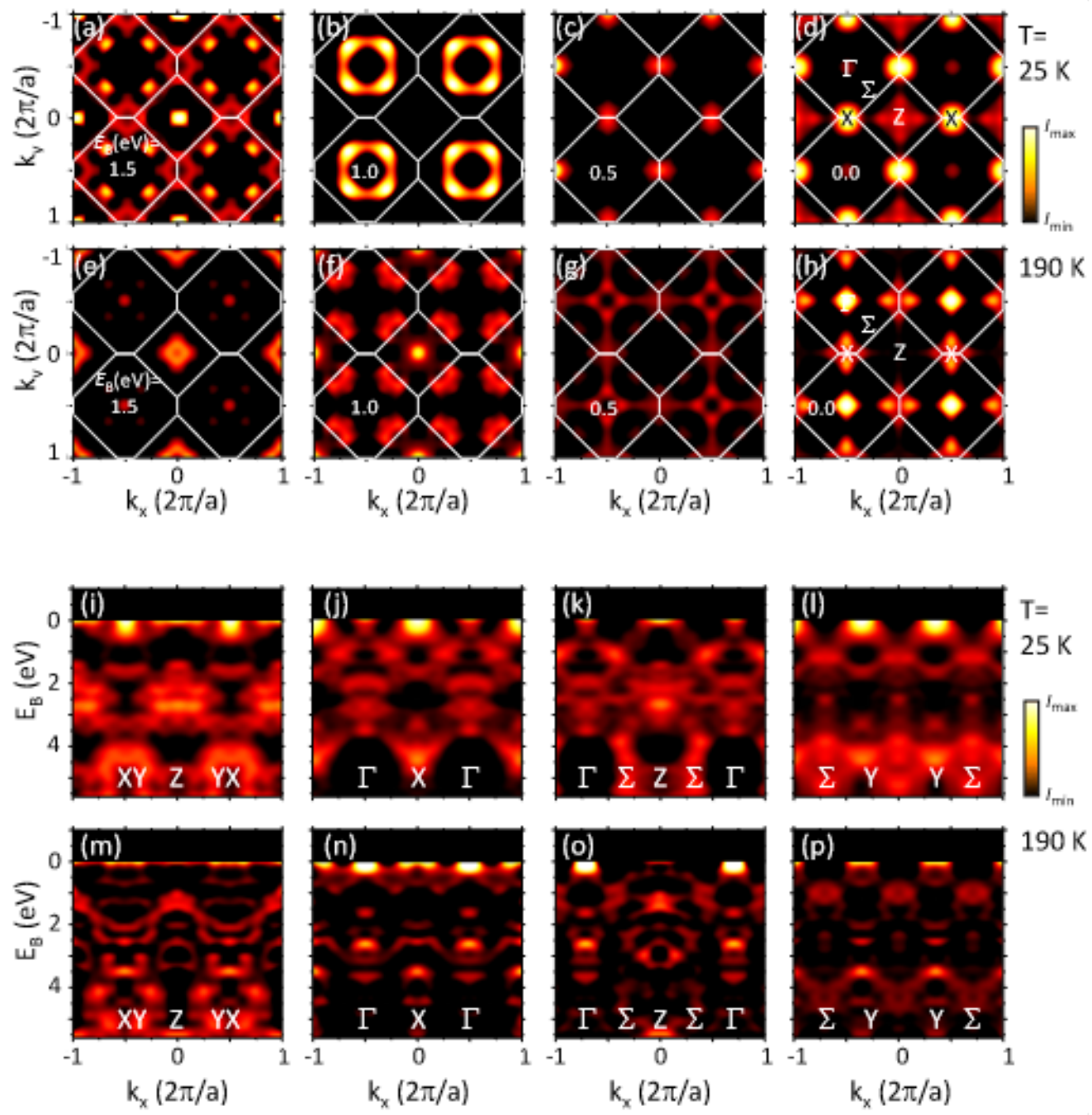

FIG. 8. (a-h) $Eu(Pd_{1-x}Au_x)Si_2$ with $x = 0.03$. Constant energy maps of the photoemission intensity $I(E_B, k_x, k_y)$ for the indicated binding energies $E_B$ measured at 25 K (a-d) and 190 K (e-h). The photon energy is 3.4 keV. The photoemission intensity has been symmetrized according to the crystal symmetry. (i-p) Binding energy versus parallel momentum sections of the photoemission data array along the indicated high symmetry directions measured at 25 K (i-l) and 190 K (m-p). The photoemission intensity is normalized on the valence band energy distribution curve shown in Fig. 3(b) and color coded on the indicated linear orange-hot scale.

To study the influence of the valence transition region on the valence band in more detail, we measured the temperature dependence of the integrated valence band intensity. Figure 7 shows the result, indicating a significant decrease of the Eu $4f$ intensity peak at $E_B = 0.5$ eV related to the $Eu^{2+}$ ionization state with decreasing temperature. The intensity at $E_B = 8$ eV, related to the Eu $4f$ states of $Eu^{3+}$ increases instead. This behavior is qualitatively similar as the results for $EuPd_2(Si_{1-x}Ge_x)_2$ shown in Fig. 3. In contrast to $EuPd_2(Si_{1-x}Ge_x)_2$ the Eu $4f$ states of $Eu^{3+}$ at $E_B = 8$ eV show the typical $4f$ multiplet of the mixed valence state in the high temperature phase, too. This indicates that the valence transition to the high temperature state is not as complete as for $EuPd_2(Si_{1-x}Ge_x)_2$.

The ratio of the intensity peak at $E_B = 0.5$ eV and -3.5 eV, shown in the inset of Fig. 7, reveals a transition temperature at 75 K in good agreement with previously published results[3].

Figure 8 shows the momentum-dependent photoemission intensities for $Eu(Pd_{1-x}Au_x)Si_2$ with $x = 0.05$. At 25 K, the constant energy intensity map $I(E_B, k_x, k_y)$ at the Fermi level [Fig. 8](d) in the Γ-Σ-X plane shows high cicular shaped intensities at the X-points. These high intensity areas shrink at $E_B = 0.5$ eV and at $E_B = 1$ eV circles of high intensity appear around the Γ points. At 190 K above $T_V$, the high photoemission intensity at the X-points becomes lower and elongated along the Γ-X direction [Fig. 8](h)]. Instead a high intensity now occurs at the Γ points. At $E_B = 1$ eV [Fig. 8](f)] the circles of high intensity are much larger and the areas of high intensity are shifted to the Σ-points.

At low temperature, the band dispersion patterns show the electron bands near the X-points with an electron-like parabolic dispersion and a maximum binding energy of 0.2 eV [see Fig. 4(i,j,l)]. At 190 K, this band has shifted to the Fermi level and is just barely visible. [see Fig. 4(m,o,p)]. Therefore, the behavior of these electron-like bands is similar to the case of $EuPd_2(Si_{1-x}Ge_x)_2$ with $x = 0.05$.

In contrast to the case of $EuPd_2(Si_{1-x}Ge_x)_2$, a second electron-like band crosses the Fermi level at the Γ-point both at low and high temperature. In addition, the band with mean value at $E_B = 1$ eV appearing in the Γ-X direction [see Fig. 4(j)] has a width of 1.5 eV, whereas it is 1 eV in the case of $EuPd_2(Si_{1-x}Ge_x)_2$ with $x = 0.05$.

These results indicate that the substitution of Pd by Au does not act just like negative pressure but also changes the electronic structure in a more drastically. This can be explained by the fact that Au is not isoelectric to Pd. Instead, the Au atom adds one electron more to the valence band as compared to Pd.

## V. DISCUSSION

Figure 9(a) illustrates the typical temperature-versus-pressure phase diagram for the valence transition compounds[28]. In this context $T_V$ represents the temperature at which the first-order valence transition temperature occurs between the high-temperature divalent Eu state and the low-temperature mixed-valent Eu2+/Eu3+ phase. In the case of the $EuPd_2Si_2$ system, previously published data have been collated from Refs.[3,5,17]. The valence transition temperature decreases with increasing substitution $x$ in $EuPd_2(Si_{1-x}Ge_x)_2$ and $Eu(Pd_{1-x}Au_x)Si_2$, whereby the substitution of atoms by larger ones acts in a manner analogous to that of negative pressure. The substitution of large amounts of Ge or Au drives the system into a low-temperature antiferromagnetic state with a Néel temperature of 30 K. In this case, the valence transition is suppressed, and the system remains in the magnetic divalent state even at very low temperatures[3].

In order to include the pressure-dependent results for $T_V(p)$, as reported in Ref.[5], in the phase diagram shown in Figure 9(a), the pressure scales for the compounds $EuPd_2(Si_{1-x}Ge_x)_2$ with $x = 0.05$ and $x = 0.1$ were adapted such that the corresponding zero pressure values align with the generic $T_V(x)$ transition line. Moreover, the pressure scales have been modified in a way that ensures the transition width of $T_V(p)$ aligns with that of $T_V(x)$. As pressure is reduced and the substitution concentration is increased, the transition width of both $T_V(p)$ and $T_V(x)$ decreases, as reported in Ref.[5]. The extrapolation to negative pressure or larger concentration, respectively, yields a critical concentration of $x_V = 0.13$.

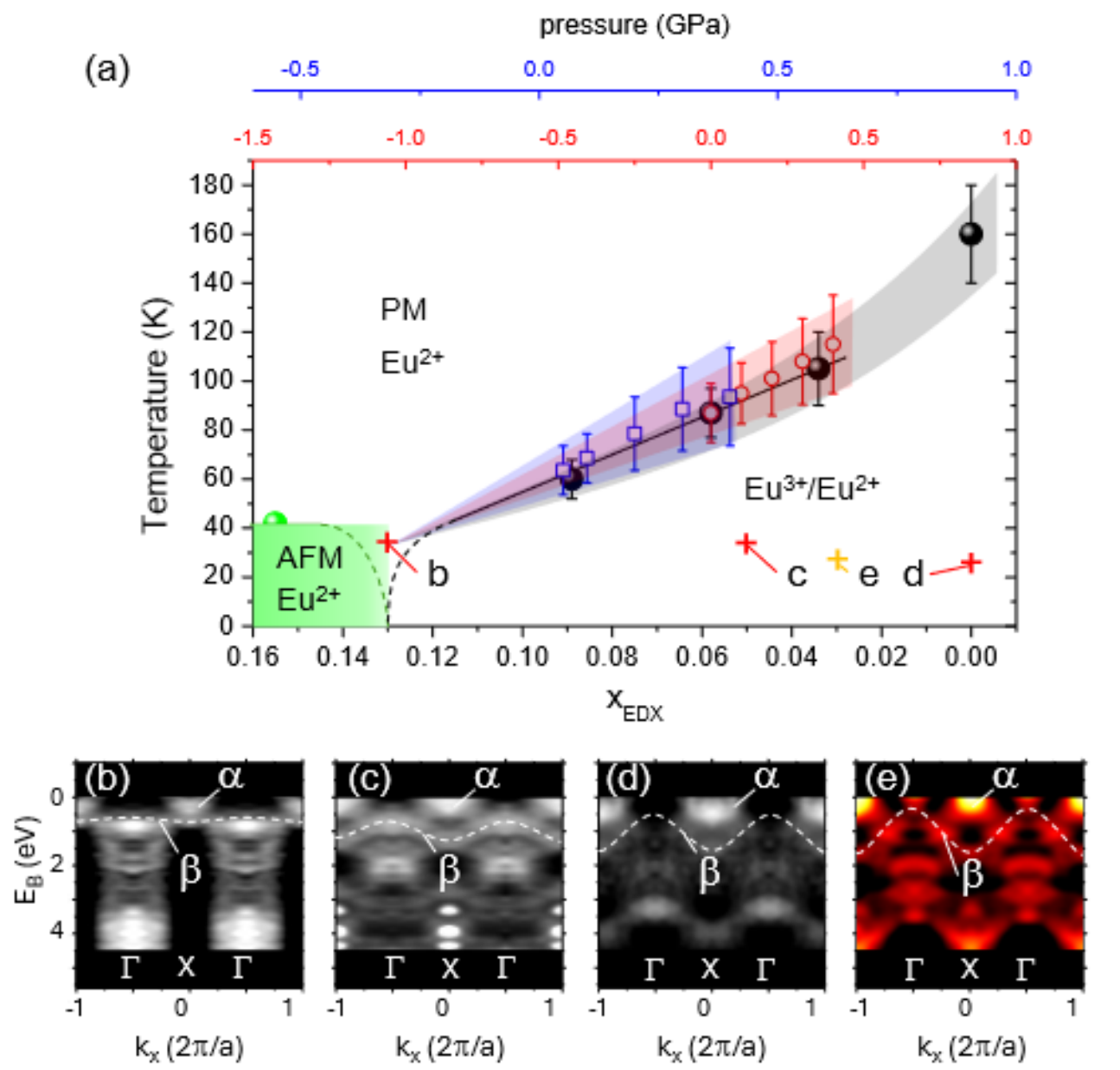

FIG. 9. (a) Phase diagram of the $EuPd_2(Si_{1-x}Ge_x)_2$ system in the vicinity of the critical endpoint. Experimental data at zero pressure from Ref.[17] (black dots) are plotted as a function of $x$. For the pressure-dependent data from Ref.[5] for $x = 0.05$ (red diamonds) and $x = 0.1$ (blue circles) the pressure scales were adapted such that $T_V$ matches the concentration-dependent data. (b) Band dispersion along the Γ-X high symmetry direction for $x = 0.13$ at $T < T_V$. (c,d) Similar data for $x = 0.05$ and $x = 0$, respectively. The data for $x = 0$ is taken from Ref.[29]. (e) Band dispersion along the Γ-X high symmetry direction for $Eu(Pd_{1-x}Au_x)Si_2$ with $x = 0.03$.

The concentration, $x_V$, or the pressure, $p_V$, is indicative of the parameter at which the first-order valence transition becomes apparent. In accordance with the findings presented in Ref.[28], there is a possibility that the antiferromagnetic state may undergo a discontinuous transition with a sharp valence crossover and the emergence of quantum critical behaviour in its vicinity[38,39]. This was considered to be realised in $EuCu_2Ge_2$ under pressure[40,41].

In order to examine the impact of negative pressure on the electronic valence band structure in $EuPd_2Si_2$, the low-temperature photoemission data along the high-symmetry Γ-X direction is compared in Fig. 9(b-e). The increasing substitution of Si by isoelectric Ge results in a shift of the electron-like $\alpha$-band, centered at the X-point, to lower binding energy. In the case of the critical concentration $x = 0.13$ the band is split into two subbands, which is indicative of the onset of exchange splitting. The $\beta$-band, centered at a binding energy of 1 eV, displays a reduction in width of dispersion with increasing Ge concentration. This decrease in width may be attributed to a reduction in hybridization with increasing atomic distances and a corresponding localization of the electronic states. At the critical concentration, the band becomes almost flat and reduced in dimensions, which may be attributed to the onset of strong fluctuations in the vicinity of the critical end point.

In contrast, the substitution of Pd by Au results in the emergence of an additional electron-like band occurring at the Fermi level at the Γ-point, which is not observed in $EuPd_2(Si_{1-x}Ge_x)_2$. Moreover, the $\beta$-band exhibits a greater width than that observed in $EuPd_2(Si_{1-x}Ge_x)_2$. These discrepancies are likely attributable to the Au valence states providing an additional electron per atom to the valence band.

## VI. SUMMARY

Hard X-ray angle-resolved photoemission spectroscopy is employed to measure the electronic bulk properties of the valence transition system $EuPd_2Si_2$ with atomic Au and Ge substitutions on the Pd and Si sites, respectively. Notable modifications in the valence band states are evident at a the valence transition, where the Eu ions undergo a temperature-induced valence crossover from a magnetic $Eu^{2+}$ state to a low-temperature valence-fluctuating state. A systematic alteration of the band structure is observed for the isoelectric substitution of Si by Ge in $EuPd_2(Si_{1-x}Ge_x)_2$. In contrast the Pd substitution by Au in $Eu(Pd_{1-x}Au_x)Si_2$ results in the emergence of additional band states near the Fermi level. Previously published pressure-dependent data for $EuPd_2(Si_{1-x}Ge_x)_2$ indicate the Ge concentration of $x = 0.13$ is situated in proximity to the critical end point. In this case, a splitting of conduction band states and the emergence of flat bands with one-dimensional character have been observed.

## ACKNOWLEDGMENTS

This work was funded by the Deutsche Forschungsgemeinschaft (DFG, German Research Foundation), grant no. TRR288–422213477 (projects B04, A03, and A05), and by the BMBF (projects 05K22UM2 and 05K22UM4). Funding for the instrument by the Federal Ministry of Education and Research (BMBF) under framework program ErUM is gratefully acknowledged. We acknowledge DESY (Hamburg, Germany), a member of the Helmholtz Association HGF, for the provision of experimental facilities. Parts of this research were carried out at PETRA III using beamline P22. O.F. acknowledges funding by TopDyn. H.J.E. thanks Denys Vyalikh for fruitful discussions.

---

* elmers@uni-mainz.de